\documentclass{aa}  
\usepackage{graphicx}
\usepackage{txfonts}
\usepackage{pslatex}
\usepackage{natbib}
\usepackage{multirow}
\begin{document} 

   \title{Direct probe of the inner accretion flow\\around the supermassive black hole in NGC 2617
   }

   \author{M. Giustini
          \inst{1,2}
         \and
         E. Costantini\inst{1}
         \and
                  B. De Marco\inst{3}
                  \and
                 J. Svoboda\inst{4}
                  \and
                   S. E. Motta\inst{5}
                  \and
                  D. Proga\inst{6}
                  \and
                  R. Saxton\inst{2}
        \and
        C. Ferrigno\inst{7} 
       \and
              A. L. Longinotti\inst{8} 
       \and
         G. Miniutti\inst{9}  
                  \and
                  D. Grupe\inst{10}
                  \and
                  S. Mathur\inst{11}
                  \and
                  B. J. Shappee\inst{12, 13}
                  \and
                  J. L. Prieto\inst{14,15}
                  \and
                   K. Stanek\inst{11}
          }

    \institute{SRON Netherlands Institute for Space Research, Sorbonnelaan 2, 3584 CA Utrecht, the Netherlands\\
              \email{m.giustini@sron.nl}
         \and
             XMM-Newton Science Operations Centre, ESA/ESAC, Apartado 78, 28692 Villanueva de la Ca\~nada, Madrid, Spain
       \and
       Max-Planck-Institut f\"ur Extraterrestrische Physik, Giessenbachstrasse 1, D-85748, Garching, Germany
       \and
              Astronomical Institute, Academy of Sciences, Bocn\'i II 1401, CZ-14100 Prague, Czech Republic
        \and
        University of Oxford, Department of Physics, Astrophysics, Denys Wilkinson Building, Keble Road, Oxford OX1 3RH, UK
       \and
       Department of Physics and Astronomy, University of Nevada, Las Vegas, NV 89154, USA
       \and
       ISDC, University of Geneva, Chemin d'\'Ecogia 16, CH-1290 Versoix, Switzerland
        \and
        Catedr\'atica CONACYT - Instituto Nacional de Astrof\'isica, \'Optica y Electr\'onica, Luis E. Erro 1, Tonantzintla, Puebla, C.P. 72840, M\'exico
        \and
        Centro de Astrobiolog\'ia (CSIC-INTA), Dep. de Astrof\'isica; ESAC, PO Box 78, Villanueva de la Ca\~nada, E-28692 Madrid, Spain
        \and
        Department of Earth and Space Science, Morehead State University, 235 Martindale Dr., Morehead, KY 40351, USA
        \and
        Astronomy Department, Ohio State University, Columbus, OH 43210, USA
        \and
        Carnegie Observatories, 813 Santa Barbara Street, Pasadena, CA 91101, USA
        \and
        Hubble, Carnegie-Princeton Fellow
        \and
         N\'ucleo de Astronom\'ia de la Facultad de Ingenier\'ia, Universidad Diego Portales, Av. Ej\`ercito 441, Santiago, Chile
         \and
         Millennium Institute of Astrophysics, Santiago, Chile
        }

 
  \abstract
{}   { NGC 2617 is a nearby ($z\sim 0.01$) active galaxy that recently switched from being a Seyfert 1.8 to be a Seyfert 1.0.
At the same time, it underwent a strong increase of X-ray flux by one order of magnitude with respect to archival measurements.  
We characterise the X-ray spectral and timing properties of NGC 2617 with the aim of studying the physics of a changing-look active galactic nucleus (AGN). }
   { We performed a comprehensive timing and spectral analysis of two XMM-Newton pointed observations spaced by one month, complemented by archival quasi-simultaneous INTEGRAL observations.}
   {We found that, to the first order, NGC 2617 looks like a type 1 AGN in the X-ray band and, with the addition of a modest reflection component,  its continuum can be modelled well either with a power law plus a phenomenological blackbody, a partially covered power law, or a double Comptonisation model. 
   Independent of the continuum adopted, in all three cases a column density of a few $10^{23}$ cm$^{-2}$ of neutral gas covering 20-40\% of the continuum source is required by the data. Most interestingly, absorption structures due to highly ionised iron have been detected in both observations with a redshift of about $0.1c$ with respect to the systemic redshift of the host galaxy.}
   {The redshifted absorber can be ascribed to a failed wind/aborted jets component, to gravitational redshift effects, and/or to matter directly falling towards the central supermassive black hole.
In either case, we are probing the innermost accretion flow around the central supermassive black hole of NGC 2617 and might be even watching matter in a direct inflow towards the black hole itself.}

\titlerunning{Direct probe of the inner accretion flow around a supermassive black hole}
\authorrunning{M. Giustini et al.}
   \maketitle


\section{Introduction}
Active Galactic Nuclei (AGN) are among the most luminous objects in the Universe, and are powered by mass accretion onto supermassive black holes (SMBH) residing at the centre of galaxies \citep{1964ApJ...140..796S, 1969Natur.223..690L, 1984ARA&A..22..471R, 1995ARA&A..33..581K}.

In the optical band, AGN can appear as unobscured with a strong continuum emission and the superposition of both broad and narrow emission lines coming from the so-called broad and narrow line regions (type~1 AGN), respectively. Alternatively, they can show up to very high levels of obscuration with the continuum and broad emission lines strongly suppressed and with only narrow emission lines visible (type~2 AGN). 
According to the scenario that currently best explains the AGN phenomenology, to the first order we are observing the same kind of object at different inclination angles: an anisotropic cold absorber located at parsec scales (the so-called ``torus'') is blocking the view of the central engine in the case of type~2 AGN, leaving it more and more unblocked towards type~1 AGN (Antonucci 1993).  
The real situation is more complex, where the cold absorber is probably not homogeneous but is patchy and clumpy, and is not static but rather part of the inflow/outflow of matter around the SMBH \citep[see e.g.][and references therein]{2012AdAst2012E..17B}.  

Several AGN have also been observed to change class in different observations. This phenomenon could either be extrinsic, related to the clumpy nature of the absorbers \citep[i.e. clouds crossing the line of sight could cause a temporary occultation of the continuum source; see e.g.][]{2005ApJ...623L..93R, 2014Sci...345...64K} or intrinsic to the AGN central engine evolution; below a certain mass accretion rate, the broad line region is not yet expected to  be formed, therefore broad emission lines are not expected, so no type~1 AGN should be observed \citep[e.g. ][]{2000ApJ...530L..65N, 2014MNRAS.438.3340E}.

While overwhelming evidence for matter outflowing from the inner regions of AGN has recently emerged as blueshifted UV and/or X-ray absorption features \citep[see e.g.][]{2003ARA&A..41..117C, 2010SSRv..157..265C, 2010A&A...521A..57T, 2013MNRAS.430...60G}, there is general lack of strong evidence of matter inflowing towards the central SMBH \citep[see][]{2006AN....327.1012C}.
In fact, redshifted X-ray absorption has been claimed in several AGN, although whenever such sources have been re-observed, the associated redshifted X-ray absorption has never been detected again \citep[e.g.][]{1999ApJ...523L..17N, 2005A&A...442..461D, 2005ApJ...633L..81R, 2007MNRAS.374..237L, 2010ApJ...717..209C}.

On April 26 2013, Shappee and collaborators reported the discovery of a substantial optical brightening of the nearby \citep[$z=0.0142$, luminosity distance $D_L=62$ Mpc;][]{2003A&A...412...57P} Seyfert 1.8 galaxy NGC 2617 (J2000 Equatorial coordinates RA=08h35m38.79s, -04h05m17.6s) along with its transition to Seyfert 1.0\footnote{\url{http://www.astronomerstelegram.org/?read=5010}}. The day after, XMM-Newton \citep{2001A&A...365L...1J} pointed NGC 2617 for 66 ks as a Director Discretionary Time observation (OBS1 from now on), and SWIFT started monitoring the source on a (almost) daily basis. After about one month, NGC 2617 was found by SWIFT-XRT to have increased its $0.3-10$ keV flux by a factor of two and a second 35 ks XMM-Newton Target of Opportunity observation was triggered on May 24 2013 (OBS2 from now on). The results of the extensive multiwavelength campaign performed on NGC 2617 are reported in \citet{2014ApJ...788...48S}. 

Here we focus on the analysis of the two high quality, high signal-to-noise ratio (S/N) XMM-Newton observations. 
This article is structured as follows: Section \ref{Section:OBS} contains information about the observations and data reduction; Section~\ref{Section:DATA} presents the data analysis; and in Section~\ref{Section:DISCU} we discuss the results, which are then summarised in Section~\ref{Section:CONCLU}.

A cosmology with $H_0 = 70$ km s$^{-1}$ Mpc$^{-1}$, $q_0=0$, and $\Omega_{\Lambda}=0.73$ is adopted throughout the paper.

\section{Observations and data reduction}\label{Section:OBS}

\begin{table*}
\caption{Log of the EPIC-pn observations\label{T_OBS}}
\centering
\begin{tabular}{cccccc}
\hline\hline
OBSID &  Date           & t$_{exp}$ & t$_{net}$ & ct s$^{-1}$ & ct s$^{-1}$ \\
(1) & (2) & (3) & (4) & (5) & (6) \\
\hline
0701981601      & 27/04/2013 & 65.7 & 46.6 & 13.68$\pm{0.02}$ & 6.80$\pm{0.01}$\\ 
0701981901      & 24/05/2013 & 34.6 & 14.8 & 32.27$\pm{0.05}$ & 16.24$\pm{0.03}$ \\
\hline
\end{tabular}
\tablefoot{(1) Observation ID; (2) Date of observation (dd/mm/yyyy); (3) Exposure time (ks); (4) Net exposure time after accounting for deadtime effects and flaring background removal (ks); (5) count rate in the 0.3-10 keV band; (6) count rate in the $0.3-10$ keV band after excising the central 5 arcseconds of the source extraction region and retaining single events only.}
\end{table*}

XMM-Newton observed NGC 2617 twice, at the end of April 2013 ($\sim 66$ ks, OBS1) and at the end of May 2013 ($\sim 35$ ks, OBS2), for a total of $\sim 100$ ks. Details of the European Photon Imaging Camera (EPIC) pn \citep{2001A&A...365L..18S} observations, which have the highest S/N that we use on this work, are reported in Table~\ref{T_OBS}.
Both EPIC-pn observations used the thin optical filter and were taken in Small Window mode (dead time of $29\%$, which gives an effective exposure time 46.6 and 24.6 ks for OBS1 and OBS2, respectively). 
The Observation Data Files (ODF) were processed with the Science Analysis System (SAS)  v.14.0.0 using calibration files generated in November 2014. 
Data were reprocessed using standard SAS analysis threads, using the task \texttt{epproc} to concatenate the raw events.

We then searched for strong flaring background time intervals during our observations to discard these intervals and retain only time intervals dominated by the source emission. To this end, after the inspection of a light curve of single events (PATTERN==0) of the whole field of view taken at 10 keV $< E <$ 12 keV, with quality flag \#XMMEA$\_$EP, the whole exposure was retained for the scientific analysis in the case of OBS1. The recommended threshold of 0.4 ct s$^{-1}$ was applied to filter the event file in the OBS2 case, reducing the effective exposure time from 24.6 to 14.8 ks.  

Source and background events were extracted from circular regions with a radius of 40'', retaining single and double pattern events. All the events were used for the timing analysis, while the spectral analysis was focused on single events only, which present the best energy resolution. Ancillary response file and response matrix at the source position were generated with the \texttt{arfgen} and \texttt{rmfgen} tasks.
The presence of pile-up was checked with the \texttt{epatplot} task, and slight deviations (excess of singles and deficit of doubles) from the predicted event pattern distribution were found in both observations. In particular, an excess of single and a deficit of double events was observed. Since these deviations are observed in the soft band we ascribe them to X-ray loading rather than to pile-up. X-ray loading occurs for very bright sources, which are able to contaminate the offset map with extra X-ray photons; the dark current in each CCD is then over-subtracted during the reprocessing of the data, and the resulting spectrum looks softer than the actual spectrum. We therefore reprocessed the raw data again applying the correction for X-ray loading by passing the parameter \texttt{runepxrlcorr=yes} to the \texttt{epproc} task, and for pile-up by excising the central five arcseconds of the point spread function of the source extraction region. A comparison of the uncorrected and corrected spectra indeed revealed slight deviations of the order of 2\% at energies softer than $\sim 2$ keV. We checked that these deviations do not affect the main results of our analysis.

The MOS \citep{2001A&A...365L..27T} data were heavily piled-up and therefore not useful for a scientific analysis. 
The Reflection Grating Spectrometer \citep[RGS;][]{2001A&A...365L...7D} data were reduced using the \texttt{rgsproc} task, discarding the cool pixels. 
The Optical Monitor \citep[OM; ][]{2001A&A...365L..36M} data, taken with the UVW1, UVM2, and UVW2 filters for OBS1 and with the UVW1 filter only in the case of OBS2, were reduced using the \texttt{omichain} task.

NGC 2617 was also observed by INTEGRAL \citep{2003A&A...411L...1W}.
Using the Offline Analysis Software (OSA) version 10.2 distributed by the ISDC \citep{isdc}, we analysed all the available INTEGRAL data from 2013-04-22 11:56:17 to 2014-11-24 00:38:24 UTC whenever the source was less than 10 degrees off-axis. We extracted the IBIS image \citep{ibis} in the energy bands 20-40 keV and 40-100 keV using the lower energy detector ISGRI \citep{isgri}: NGC 2617 was detected with a confidence level of 15$\sigma$ and 11$\sigma$. 
The vignetted and dead-time corrected exposure time was 437.6 ks. We extracted an ISGRI light curve with bins of 2 ks and found that the source flux level showed undetectable variability over the analysed time period.

\section{Data analysis}\label{Section:DATA}

Data analysis was performed using \textsc{Heasoft v.6.16} \citep{1995ASPC...77..367B}, \textsc{Xspec v.12.9.0i} \citep{1996ASPC..101...17A}, and \textsc{SPEX v.2.05.04} \citep{1996uxsa.conf..411K}. Error bars plotted in figures and quoted in the text are at the $1\sigma$ confidence level if not otherwise stated.

\subsection{Properties of X-ray time variability\label{SEC:TIMING}}
\begin{figure*}
   \centering
  \includegraphics[width=8. cm]{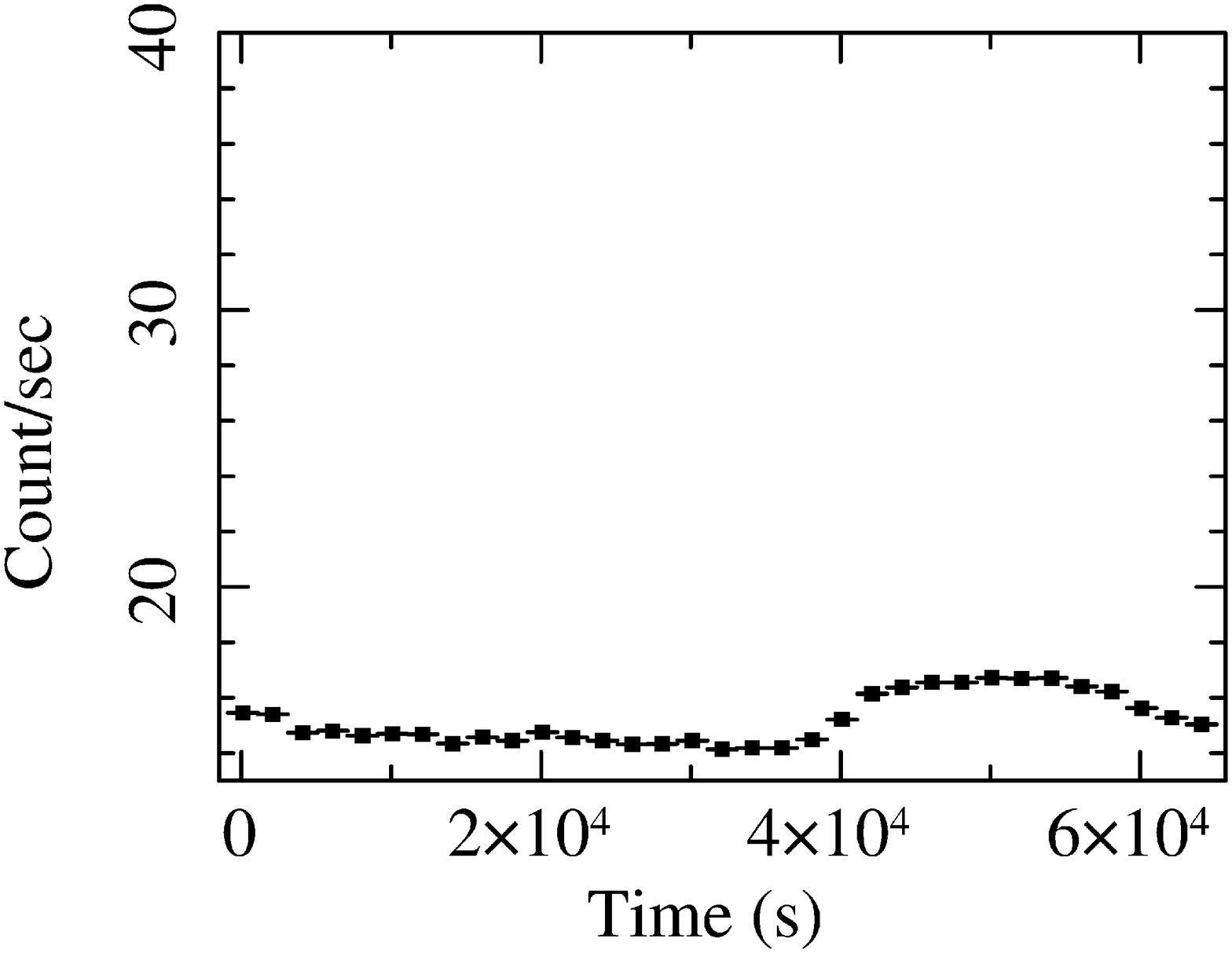}
  \includegraphics[width=5. cm]{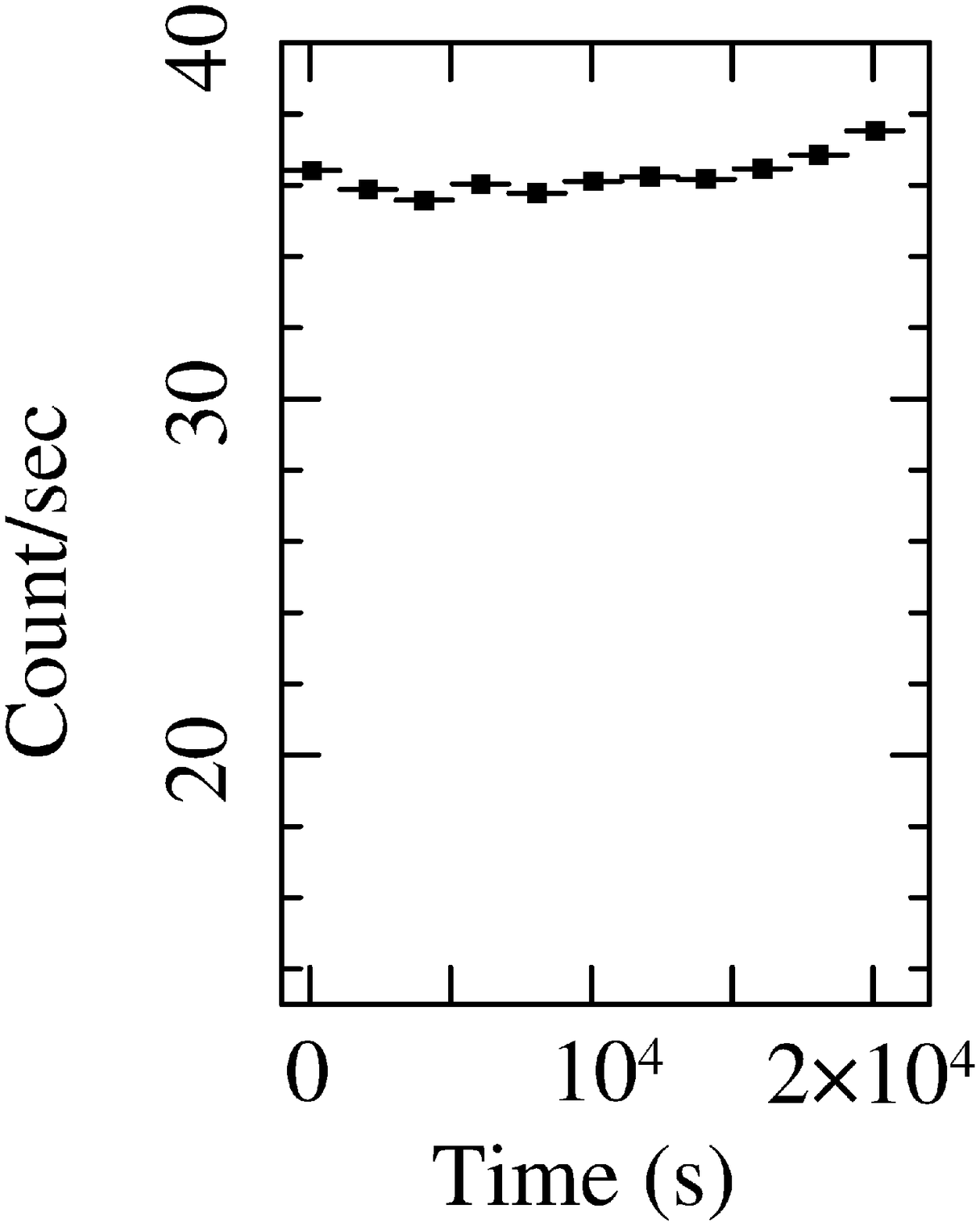}
   \caption{OBS1 (left) and OBS2 (right) EPIC-pn light curve of NGC 2617 extracted in the $0.3-10$ keV band and binned to 2 ks. 
   The OBS2 light curve is truncated at the end of the observation, when strong background flaring counts dominate over the source counts (see the text for details). }
              \label{FIGURE1}
    \end{figure*}

\begin{figure*}
   \centering
  \includegraphics[width=13.5cm]{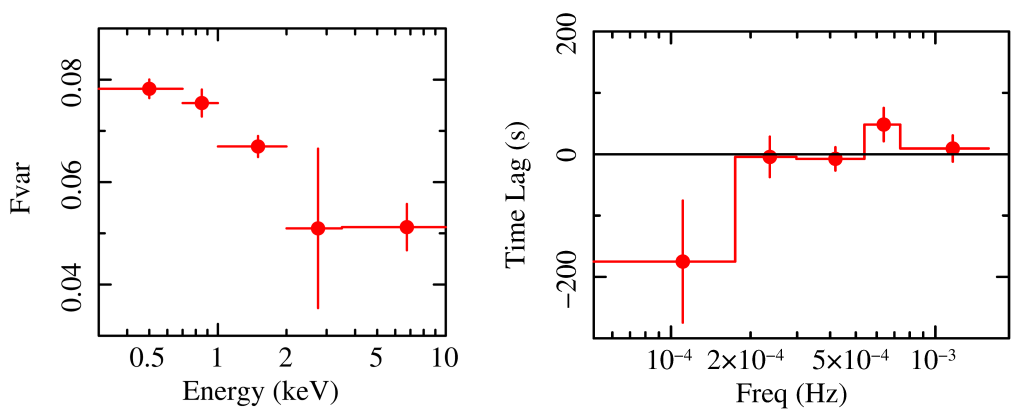}
     \caption{Left panel: Fvar of NGC 2617, computed on timescales of 3-20 ks. Right panel: lag-frequency spectrum between the $0.3-1$ and $2-10$ keV band.}
              \label{FIGURE2}
    \end{figure*}

We analysed the X-ray short-term time variability of NGC 2617.
The EPIC-pn light curves of NGC 2617 extracted in the $0.3-10$ keV band and binned to 2 ks are shown in Figure~\ref{FIGURE1}, where the task \texttt{epiclccorr} was used to apply both absolute (e.g. vignetting, bad pixels, and chip gaps) and relative (e.g. dead time, GTIs, and background level) time-dependent corrections. 
The source shows modest X-ray flux variations and a hardness ratio analysis revealed the absence of strong spectral variability within each observation.
We computed the short timescales (i.e. between about 3 ks and 20 ks) fractional root-mean-square variability amplitude \citep[Fvar; e.g.][]{1997ApJ...476...70N,2003MNRAS.345.1271V}, finding values of a few percent (see Fig.~\ref{FIGURE2}, left panel) with hints of a slight increase of Fvar at softer energies. These values of Fvar are typical of a source with $M_{BH}>10^7 M_{\odot}$ \citep{2012A&A...542A..83P}. This is in agreement with the NGC 2617 black hole mass estimate of $4\times 10^7 M_{\odot}$ inferred from the H$\beta$ line width and the radius-luminosity relation by \citet{2014ApJ...788...48S}. 
Nonetheless, the source flux more than doubled between the two exposures (Fig.~\ref{FIGURE1}), implying that NGC 2617 is much more variable on long timescales (of the order of days) than on short timescales (of the order of hours). 

We finally checked for the presence of time delays between X-ray energy bands. We did so by extracting light curves in the $0.3-1$ (soft) and $2-10$ keV (hard) bands from the two observations, and computing time lags in the Fourier-frequency domain \citep[e.g.][]{2013MNRAS.431.2441D, 2014A&ARv..22...72U}. We find hints of a 200 s soft band delay on timescales that are larger than $\sim 5$ ks (i.e. frequencies lower than $\sim 2\times 10^{-4}$ Hz; see Fig. 2, right panel).

\subsection{Spectral analysis: The broadband continuum\label{SEC:CONTINUUM}}
The EPIC-pn spectra were grouped with the SAS \texttt{specgroup} task to have a minimum S/N of 10 in each energy bin. 
We used the  $\chi^2$ statistics in the search for the best-fit model and for parameter errors determination \citep{1976ApJ...210..642A}, and quoted statistical errors at the 1$\sigma$ confidence level (i.e., $\Delta\chi^2=1$ for one parameter of interest). Every model included a Galactic column density $N_H^{Gal}=3.64 \times 10^{20}$ cm$^{-2}$ \citep{2005A&A...440..775K} that was modelled with \texttt{tbabs} (Wilms et al. 2016, in prep.) within \textsc{Xspec}, and with \texttt{hot} within \textsc{SPEX}. In the latter case the temperature of the absorbing gas was set equal to 0.5 eV to mimic a cold phase. 

We focused the analysis on the 0.3$-$10 keV band and fit the OBS1 and OBS2 spectra simultaneously, thereby allowing for the minimum number of parameters to vary between the two epochs of observation, i.e. constant parameters were tied to a common value in OBS1 and OBS2.

     \begin{figure}
   \centering
  \includegraphics[width=6. cm,angle=-90]{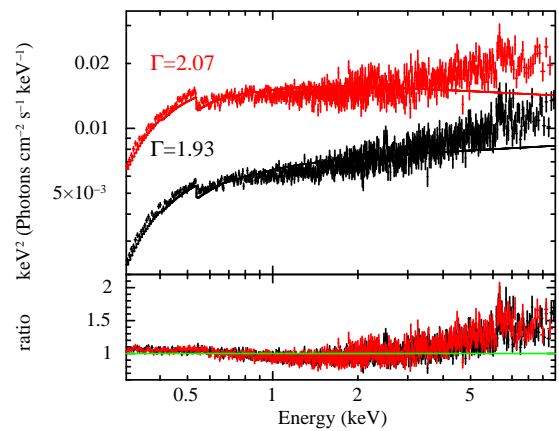}
   \caption{Top panel: OBS1 (black, lower flux) and OBS2 (red, higher flux) 0.3-10 keV EPIC-pn spectra unfolded against a power law emission plus Galactic absorption model. Bottom panel: corresponding data/model ratio. Data were visually rebinned to a significance of 10$\sigma$ (Xspec command \texttt{setplot rebin}).}
              \label{FIGURE3}%
    \end{figure}
    
The top panel of Figure~\ref{FIGURE3} shows the OBS1 and OBS2 spectra unfolded against a power law emission model, where the slope and normalisation were allowed to change between the epochs. The bottom panel shows the corresponding data/model ratio. 
The fit is clearly unacceptable, with $\chi^2/\nu=3932/1339$. 
The X-ray spectrum of NGC 2617 is evidently complex, showing important departures from a simple power law both at soft and especially at high energies.

   \begin{figure}
   \centering
  \includegraphics[width=3.65 cm,angle=-90]{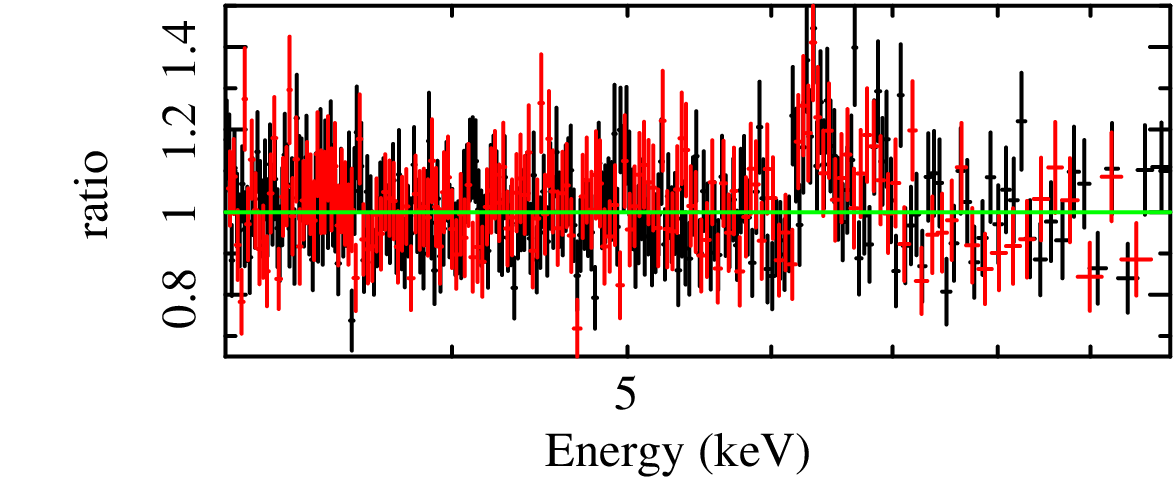}
  \includegraphics[width=3.65 cm,angle=-90]{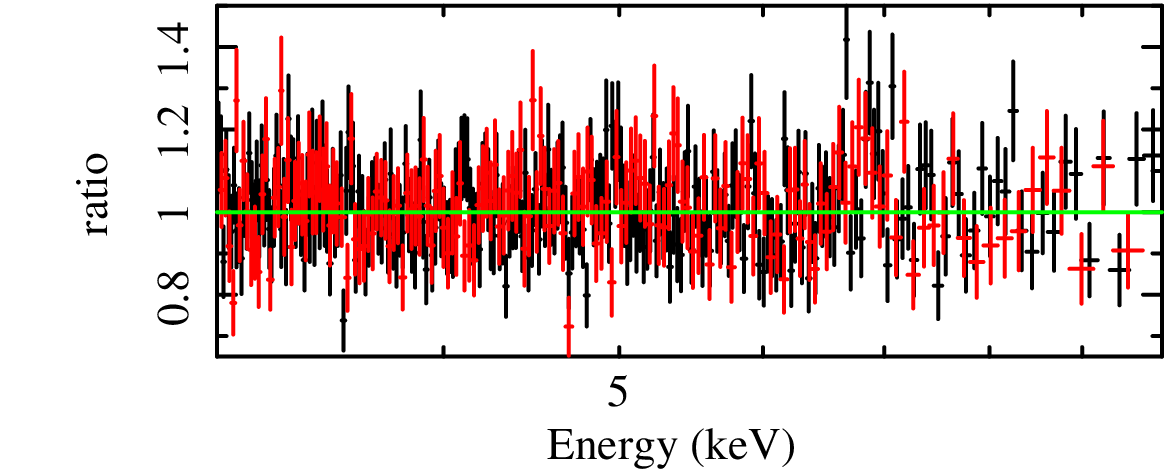}
  \includegraphics[width=3.75 cm,angle=-90]{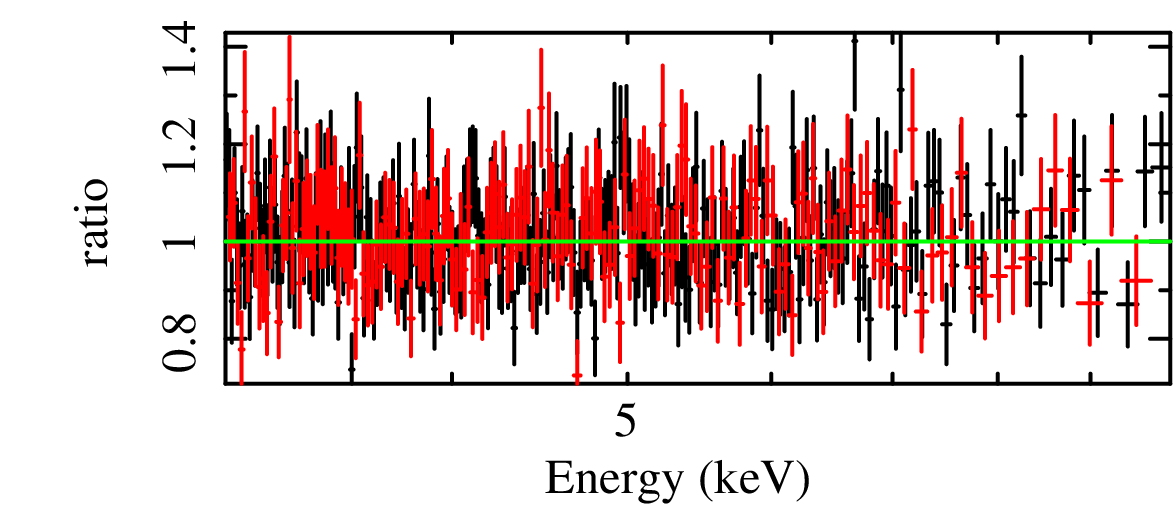}
  \includegraphics[width=3.65 cm,angle=-90]{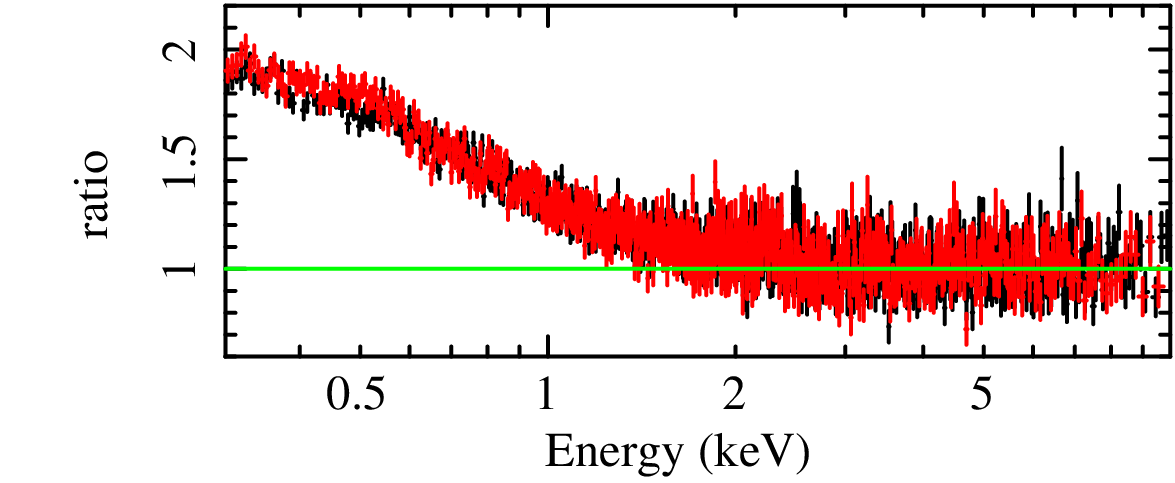}
   \caption{Top panel: EPIC-pn spectral residuals for a power law emission plus Galactic absorption model fitted in the $3-10$ keV band, for OBS1 (black) and OBS2 (red). Middle panel: same as above after the inclusion of a Gaussian emission line. Bottom panel: data/model ratio over the full energy band.}
              \label{FIGURE4}%
    \end{figure}
    
If restricted to the $3-10$ keV band, the simple power law model gives a flat slope, $\Gamma\sim 1.6$ and $\Gamma\sim 1.7$ for OBS1 and OBS2, respectively, and a fit statistics $\chi^2/\nu=479/403$. Residuals to this model are plotted in the top panel of Figure~\ref{FIGURE4}.
 
The addition of a Gaussian emission line with the energy and width fixed between the two observations significantly improves the fit statistics ($\Delta\chi^2/\Delta\nu=56/4$).
The line centroid is found at an energy $E_{line}=6.43\pm{0.02}$ keV, which is compatible with neutral or lowly ionised iron. 
The line is unresolved, with $\sigma_{line} < 112$ eV at the 90\% c. l..
The flux in the line has changed between the two exposures, following the long-term variation of the broadband source flux. 
Indeed, the equivalent width is found to be constant within the errors, EW$=69^{+11}_{-16}$ eV and EW$=69^{+14}_{-17}$ eV for OBS1 and OBS2, respectively.
Residuals to this baseline model are plotted in the second panel of Figure~\ref{FIGURE4}. Positive residuals in the Fe K band
are still present, and adding another Gaussian emission line improves the fit by $\Delta\chi^2/\Delta\nu=16/4$. 
The line is found to have an energy $E_{line}=6.96\pm{0.03}$ keV, compatible with highly ionised iron. This line is unresolved as well, with  $\sigma_{line} < 170$ eV at the 90\% c.l.. The EW is again staying constant between the two observations, with EW$=33^{+12}_{-10}$ eV and EW$=36^{+22}_{-20}$ eV for OBS1 and OBS2, respectively. Residuals to this model are plotted in the third panel of Figure~\ref{FIGURE4}
, while the full band ratio (down to $0.3$ keV) between the data and the model are shown in the bottom panel of the same figure.
Because of the flatness of the photon indices adopted, an excess of emission with respect to the $3-10$ keV power law plus Gaussian best-fit model is apparent. 
This does not appear to be the typical `soft excess' often observed in the X-ray spectra of type 1 AGN \citep{1985MNRAS.217..105A}, but rather the effect of modelling the hard band with a flat power law ($\Gamma\sim 1.6-1.7$); indeed, considering a typical power law slope of $\Gamma=1.9-2$ (as shown in Fig. \ref{FIGURE3}) the intensity of the soft excess (if present) is reduced to 10\% of the emission predicted by the power law model in the soft X-ray band.

\subsubsection{A phenomenological model}
\begin{figure*}
   \centering
 \includegraphics[width=17. cm]{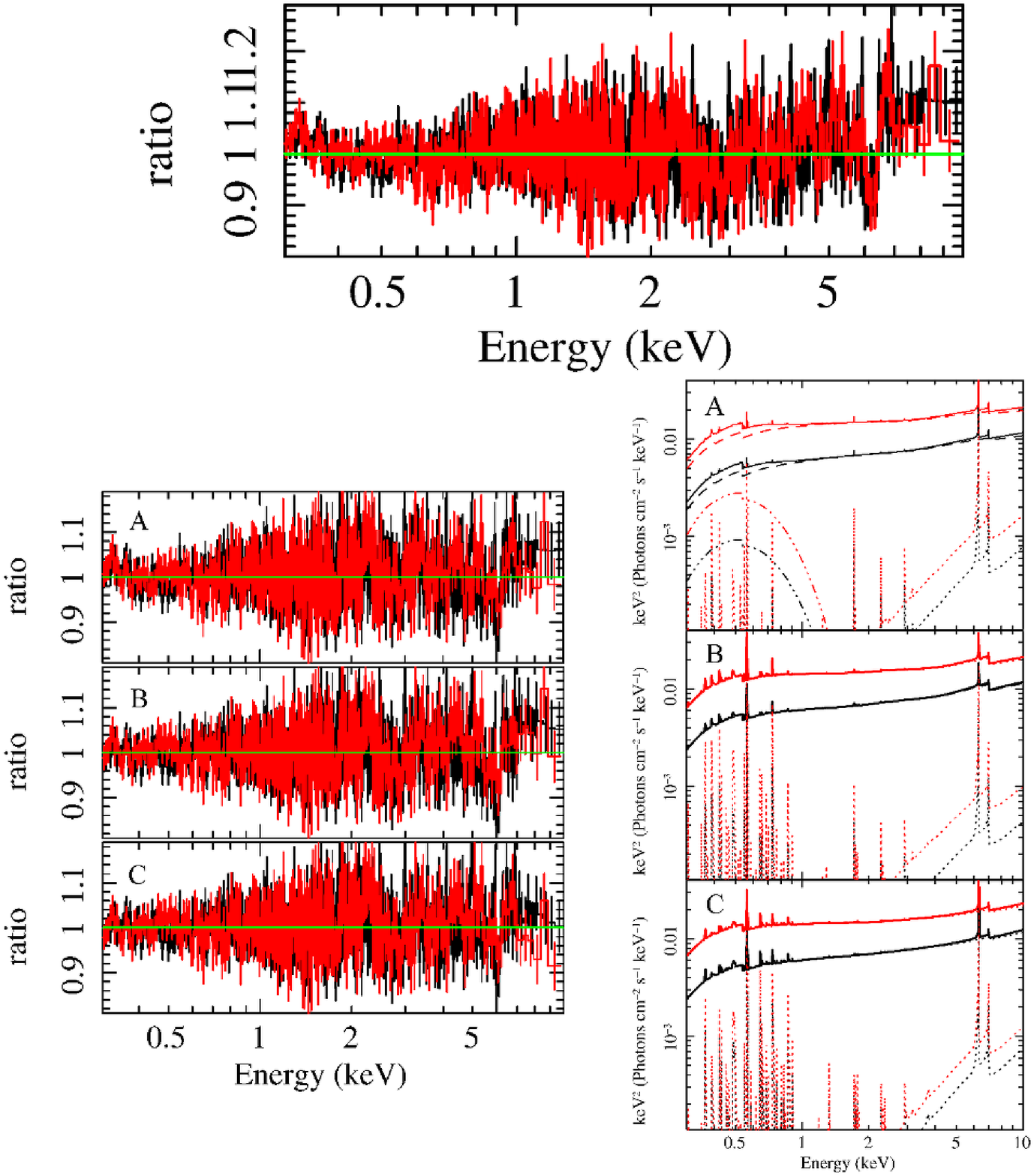}
   \caption{Top panel: EPIC-pn spectral residuals for our baseline model \texttt{xillver + tbnew\_pcf*(blackbody + power law)} fitted in the $0.3-10$ keV band, for OBS1 (black) and OBS2 (red). 
   Top left panel, Model A: spectral residuals after the addition of a cold partially covering absorber,  \texttt{xillver + tbnew\_pcf*(blackbody + power law)}. Middle left panel,  Model B: spectral residuals for a double partial covering absorber model, \texttt{xillver + tbnew\_pcf1*tbnew\_pcf2*powerlaw}. 
   Bottom left panel, Model C: spectral residuals for a double comptonisation model, \texttt{xillver + tbnew\_pcf*(optxagnf)}. 
   Data were visually rebinned to a significance of 15$\sigma$ (Xspec command \texttt{setplot rebin}).
   Top right panel: theoretical model A; middle right panel: theoretical model B; bottom right panel: theoretical model C. Black lines always refer to OBS1, and red lines to OBS2.}
              \label{FIGURE5}%
    \end{figure*}

Turning our attention on the full $0.3-10$ keV band, and noticing the slight excess of soft X-ray emission and the strong complexities in the Fe K band (Fig.~\ref{FIGURE4}), we added a blackbody emitter to fit the slight soft excess to the baseline model. We replaced the single Gaussian emission line with a self-consistent reflection component, using the \texttt{xillver} model \citep{2013ApJ...768..146G}, where the normalisation was left free to vary and the ionising continuum slopes were tied to the slope of the primary power law emission in the two different epochs.
The fit statistics for this model is $\chi^2/\nu=1537/1333$ and spectral residuals are plotted in the top central panel of Figure~\ref{FIGURE5}. 
The NGC 2617 high-energy spectrum is too hard, and the iron emission line is too weak to be reproduced by a reflection component alone, even dropping the iron abundance of the reflector to a half of solar (the minimum value allowed by the model). 
Adding a second reflection component did not improve the fit.

The addition of a layer of partially covering gas modelled with \texttt{tbnew\_pcf} (Wilms et al. in prep.) improves the fit statistics by $\Delta\chi^2/\Delta\nu=92/2$ and allows us to reproduce the continuum spectral shape of NGC 2617 well over the full energy band without the need of abundances different than solar. 
The absorber has a column $N_{H}/10^{22}\,$cm$^{-2}\sim 20$ and covers $\sim 20\%$ of the source. 
The only parameters that significantly vary between the two observations are the power law, blackbody, and reflected emission normalisations, along with the power law photon index $\Gamma$, which is found to steepen by $\sim 0.1$ going from the lower (OBS1) to the higher (OBS2) flux level.
Fit statistics for this model (\texttt{tbnew\_pcf*(blackbody + powerlaw) + xillver}, dubbed ``Model A'') are reported in the first column of Table~\ref{TABLE:FIT},  spectral residuals in the top left panel of Figure~\ref{FIGURE5}, while the corresponding theoretical model is plotted in the top right panel of the same figure.


\begin{table*}
\caption{Fit statistics for the continuum.\label{TABLE:FIT}}
\centering
\begin{tabular}{l c c | l c c | l c c}
\hline\hline
\multicolumn{3}{c}{Model A} & \multicolumn{3}{c}{Model B} & \multicolumn{3}{c}{Model C}\\
\multicolumn{3}{c}{\tiny{\texttt{xillver + tbnew\_pcf*(blackbody + power law)}}} & \multicolumn{3}{c}{\tiny{\texttt{xillver + tbnew\_pcf1*tbnew\_pcf2*powerlaw}}} & \multicolumn{3}{c}{\tiny{\texttt{xillver + tbnew\_pcf*optxagnf}}}\\
\hline
\rule{0pt}{2.75ex}  \small{ }& OBS1 & OBS2 & \small{ }& OBS1 & OBS2 & \small{ }& OBS1 & OBS2 \\
  & & &   & & &    & & \\
\rule{0pt}{2.75ex}     $^{(a)\,}\Gamma$ & 1.87$\pm{0.01}$  & 1.97$\pm{0.01}$ & $^{(a)\,}\Gamma$ & 2.028$\pm{0.007}$ & 2.146$\pm{0.008}$ & $^{(a)\,}\Gamma$ & 1.76$\pm{0.02}$ & 1.83$\pm{0.02}$ \\
\rule{0pt}{2.75ex} $^{(b)\,}$f$_{pow}$  & 3.8$\pm{0.1}$ & 8.2$^{+0.3}_{-0.2}$ & $^{(b)\,}$f$_{pow}$  & 3.9$\pm{0.2}$ & 8.4$\pm{0.5}$ & $^{(j)\,}L/L_{Edd}$ & 0.026$^{+0.195}_{-0.024}$& 0.067$^{+0.009}_{-0.008}$\\
  & & &   & & &    & & \\
\rule{0pt}{2.75ex} $^{(c)\,}kT_{bb}$  & \multicolumn{2}{c|}{111$\pm{2}$}  & $^{(e)\,}N_{H,1}$  & \multicolumn{2}{c|}{3.3$\pm{0.03}$}  & $^{(k)\,}r_{cor}$ & $>37$ & 25$^{+4}_{-10}$\\
\rule{0pt}{2.75ex}  $^{(d)\,}$f$_{bb}$  & 1.3$\pm{0.2}$ & 6.0$\pm{0.3}$ & $^{(g)\,}C_{f,1}$ & \multicolumn{2}{c|}{0.24$\pm{0.01}$}  & $^{(l)\,}kT_e$ & \multicolumn{2}{c}{270$^{+27}_{-62}$}  \\
\rule{0pt}{2.75ex}    & & &   &  & & $^{(m)\,}\tau$  &  \multicolumn{2}{c}{14$^{+8}_{-1}$}  \\
\rule{0pt}{2.75ex}  & & &   & & & $^{(n)\,}f_{pl}$  & \multicolumn{2}{c}{$0.70^{+0.11}_{-0.03}$}  \\
  & & &   & & &    & & \\
 $^{(e)\,}N_{H}$  &  \multicolumn{2}{c|}{22$\pm{4}$} & $^{(e)\,}N_{H,2}$  & \multicolumn{2}{c|}{47$^{+9}_{-7}$} & $^{(e)\,}N_{H}$ & \multicolumn{2}{c}{62$^{+51}_{-12}$} \\
$^{(g)\,}C_{f}$ & \multicolumn{2}{c|}{0.23$\pm{0.02}$} & $^{(g)\,}C_{f,2}$ & \multicolumn{2}{c|}{0.40$\pm{0.03}$} &  $^{(g)\,}C_{f,2}$ & \multicolumn{2}{c}{0.19$^{+0.05}_{-0.04}$} \\
  & & &   & & &    & & \\
 $^{(h)\,}\log\xi_R$  &  \multicolumn{2}{c|}{$<0.17$} & $^{(h)\,}\log\xi_{R}$  & \multicolumn{2}{c|}{$<0.5$} & $^{(h)\,}\log\xi_R$ & \multicolumn{2}{c}{$<1.2$} \\
 $^{(i)\,}$f$_{ref}$  & 0.9$\pm{0.3}$ & 2.3$^{+0.5}_{-0.9}$ & $^{(i)\,}$f$_{ref}$  & 0.8$\pm{0.3}$  & 2.1$\pm{0.8}$ & $^{(i)\,}$f$_{ref}$  & 1.1$^{+0.5 }_{-0.3 }$ & 2.0$^{+0.6 }_{-0.5 }$ \\
  & & &   & & &    & & \\
 $\chi^2/\nu$  & \multicolumn{2}{c|}{1445/1331} & $\chi^2/\nu$  &  \multicolumn{2}{c|}{1439/1332} & $\chi^2/\nu$  & \multicolumn{2}{c}{1415/1329}  \\
 $P_{null}$  & \multicolumn{2}{c|}{0.015} &  $P_{null}$  &  \multicolumn{2}{c|}{0.021} &  $P_{null}$  & \multicolumn{2}{c}{0.050}  \\
 
\hline
\end{tabular}
\tablefoot{Notes: \textit{(a)} Power law photon index; \textit{(b)} Observed $0.3-10$ keV power law flux in units of $10^{-11}$ erg cm$^{-2}$ s$^{-1}$ ; \textit{(c)} Blackbody peak temperature, in eV; \textit{(d)} Observed $0.3-10$ keV blackbody flux in units of $10^{-12}$ erg cm$^{-2}$ s$^{-1}$; \textit{(e)} Absorber column density in units of $10^{22}$ cm$^{-2}$; \textit{(f)} Logarithm of the absorber ionisation parameter in units of erg cm s$^{-1}$; \textit{(g)} Absorber covering factor; \textit{(h)} Logarithm of the reflector ionisation parameter in units of erg cm s$^{-1}$; \textit{(i)} Observed $0.3-10$ keV reflected flux in units of $10^{-12}$ erg cm$^{-2}$ s$^{-1}$; \textit{(j)} Eddington ratio; \textit{(k)} Coronal radius in units of gravitational radii; \textit{(l)} Electron temperature of the warm corona, in units of eV; \textit{(m)} Warm corona optical depth; \textit{(n)} Fraction of energy that inside $r_{cor}$ is dissipated in the hot, optically thin corona. Every model includes a Galactic equivalent hydrogen column density $N_H^{Gal}=3.64\times 10^{20}$ cm$^{-2}$ \citep{2005A&A...440..775K} modelled with \texttt{tbabs}, Wilms et al. 2016, in preparation.}
\end{table*}

\subsubsection{A double partial covering scenario\label{PCOV}}

A statistically acceptable representation of the $0.3-10$ keV continuum ($\chi^2/\nu=1439/1332\sim 1.08$) can also be given by a model composed of the superposition of a power law plus its reflected component, absorbed by two layers of partially covering gas (\texttt{tbnew\_pcf1*tbnew\_pcf2*(powerlaw) + xillver}, ``Model B'').

A neutral layer with $(N_{H,1}/10^{22}\,$cm$^{-2})\sim 3$ only covers $\sim 20\%$ of the source; another layer that covers about 45\% of the source has a column $(N_{H,2}/10^{22}\,$cm$^{-2})\sim 50$.
Other than the power law and reflected emission normalisation, and again the photon index that is steepening by $\sim 0.1$, the column density of the higher column density absorber is the only parameter that varies between OBS1 and OBS2.
Parameters and their errors are reported in the second column of Table~\ref{TABLE:FIT}; the corresponding model is plotted in the middle right panel of Figure~\ref{FIGURE5}, while spectral residuals are plotted in the middle left panel of Figure~\ref{FIGURE5}. 

Given their low ionisation state, the two partially covering absorbers do not imprint any atomic feature in the spectrum, instead they model the overall broadband curvature.

\subsubsection{A double Comptonisation scenario\label{COMP}}

Finally, we tested a double Comptonisation model for the continuum of NGC 2617.

We used the \texttt{optxagnf} model \citep{2012MNRAS.420.1848D}, which assumes the presence of two Comptonising regions, in addition to a radiatively efficient geometrically thin, optically thick accretion disk around the central SMBH. 
Given a mass accretion rate, which is parametrised via the Eddington ratio $L/L_{Edd}$, and a black hole spin, the disk emits as a colour-temperature corrected, blackbody from the outer radius, which we fixed to $10^5$ $r_g$, where $r_g\equiv GM_{BH}/c^2$ is the gravitational radius, down to a coronal radius $r_{cor}$ (a free parameter). Inside this radius, and up to the innermost circular orbit around the SMBH, a fraction $f_{pl}$ of the photons emitted by the disk gets Comptonised on a hot ($T=100$ keV) optically thin plasma (the ``hot corona'') and gives the hard X-ray power law with the slope $\Gamma$, while the remaining $(1 - f_{pl} )$ fraction of photons gets Comptonised in a much colder and optically thick plasma (the ``warm corona'') that is characterised by a temperature $kT_e$ and an optical depth $\tau$; the interaction of the photons with this latter plasma accounts for modelling the soft excess. We fixed the black hole mass to $4\times 10^7 M_{\odot}$ \citep{2014ApJ...788...48S}, the luminosity distance to 62 Mpc, and the black hole spin to zero for simplicity.

As in the previous cases, we found that a combination of \texttt{optxagnf + xillver} is not able to reproduce the X-ray spectrum of NGC 2617 at the highest energies probed by the EPIC-pn independent of the adopted iron abundances. Again, the introduction of a high column density layer of partially covering cold gas is able to alleviate this issue and to obtain a good fit statistics ($\chi^2/\nu=1415/1329\sim 1.06$).
Within this scenario, the warm corona has a temperature $kT_e\sim 270$ eV and an optical depth $\tau\sim 14$, and processes $30\%$ of the seed photons coming from the accretion disk. The remaining $70\%$ of the disk photons gets processed in the hot corona and gives a power law with photon index $\Gamma\sim 1.8-1.9$ depending on the epoch of observation. The steepening of $\Delta\Gamma\sim 0.1$ between OBS1 and OBS2 is still observed.

Model parameters along with their errors for \texttt{tbnew\_pcf*optxagnf + xillver} (Model C) are reported in the third column of Table~\ref{TABLE:FIT} and the theoretical model is plotted in the bottom right panel of Figure~\ref{FIGURE5}, while spectral residuals are in the bottom left panel of the same figure. 

\subsubsection{Extension at lower and higher energies\label{SEC:EXTENSION}}

Given that the double Comptonisation model also predicts emission components at optical/UV energies as the colour-corrected thermal emission from the accretion disk, we made use of the OM data as well. 
During OBS1 the $U, B, V, UVW1, UVM2,$ and $UVW2$ filters were used; during OBS2 only the $UVW1$ was used.
The $U,B,V$ filters fall in a region of the electromagnetic spectrum where there is strong contamination from the host galaxy stellar flux, so we restricted our analysis to the $UVW1, UVM2$, and $UVW2$ filters, where the contamination is minimum \citep[see e.g.][]{2015A&A...575A..22M}. 
The UV fluxes were dereddened using the \citet{1989ApJ...345..245C} algorithm and they were corrected for Fe II and Balmer continuum emission, using the NGC 5548 model studied by \citet{2015A&A...575A..22M} as our template, with the flux at 510 nm as the normalisation point.
There, the PHA files with the filter count rate were used in conjunction with the appropriate canned response matrices\footnote{\url{http://xmm2.esac.esa.int/external/xmm_sw_cal/calib/om_files.shtml}} and were fitted simultaneously to the EPIC-pn data. 

We found that the model fails to fit the OM data, independent of the parameters adopted; given the black hole mass estimate for NGC 2617, any disk emission would be much redder than the observed data points. Varying the black hole spin parameter to different values from 0 to 0.99 did not help the fit to converge, and the data are still not reproduced well by the model.

We also added data at high energies using archival \textit{INTEGRAL} IBIS/ISGRI observations from the end of April 2013 until the end of November 2014.  
We extracted a spectrum in 11 energy bins from 18 keV to 1 MeV, and found that the source could be well detected up to $\sim$100 keV. According to the latest recommendations\footnote{\url{http://www.isdc.unige.ch/integral/download/osa/doc/10.2/osa_um_ibis/node74.html}}, we first performed the spectral fitting above 22 keV, but the first bin showed unphysical residuals that are most probably due to the incorrect treatment of the pixel low-threshold values in the detector. Thus, we limited ourselves to the spectral range $28-100$ keV. 

As this is an average of many months of \textit{INTEGRAL} observations of NGC 2617, we fixed the normalisation values and the primary power law slope to have the average value between the corresponding parameters for OBS1 and OBS2. 
A statistically equivalent good fit ($\chi^2_r\sim 1.09$) was obtained for the three models considered so far. 

\begin{figure*}
   \centering
 \includegraphics[width=18. cm]{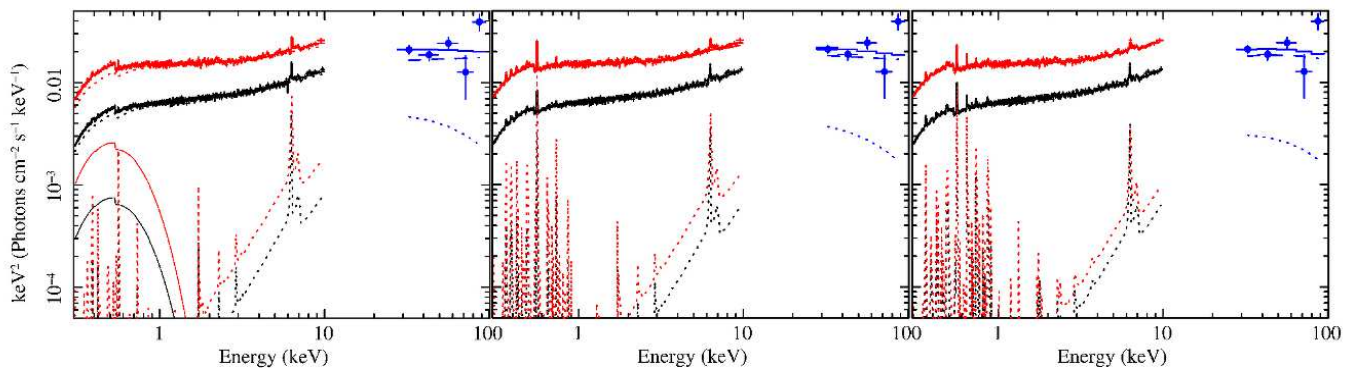}
    \caption{From left to right: unfolded XMM-Newton (black for OBS1, red for OBS2) and INTEGRAL IBIS/ISGRI (blue; data points are indicated with a circle) $0.3-100$ keV spectra for Model A, Model B, and Model C.}
              \label{FIGURE6}%
\end{figure*}

Unfolded spectra from 0.3 up to 100 keV are reported in Fig.~\ref{FIGURE6} for the three models. The differences between the three models at high energies are subtle and might require a deep observation with hard X-ray focusing telescopes to be resolved. The flux in the \textit{INTEGRAL} IBIS/ISGRI band for all the three models is $f_{28-100}\sim 5\times 10^{-11}$ erg s$^{-1}$ cm$^{-2}$.

\subsection{A redshifted absorber\label{redshift}}
 In the previous subsections we modelled the $0.3-100$ keV X-ray continuum of NGC 2617 with three statistically acceptable models.
 The inspection of the panels of Figure~\ref{FIGURE5} that report data/model ratios for the aforementioned models in the $0.3-10$ keV band reveals the presence of persistent negative residuals around 6 keV both in the OBS1 and OBS2 data sets.
 
 \subsubsection{A phenomenological model\label{redshiftedline}}
 The addition of a simple narrow (width $\sigma_{line}\equiv 0$ eV) Gaussian absorption line with the redshift fixed to the systemic value ($z=0.0142$) and centroid energy and the normalisation free to vary between the epochs improves the fit statistics by $\Delta\chi^2=21, 27,$ and $16$ for four extra degrees of freedom for Models A, B, and C, respectively ($F-$test: 99.94, 99.99, and 99.55\%). 
 
\begin{figure}
   \centering
  \includegraphics[width=9 cm]{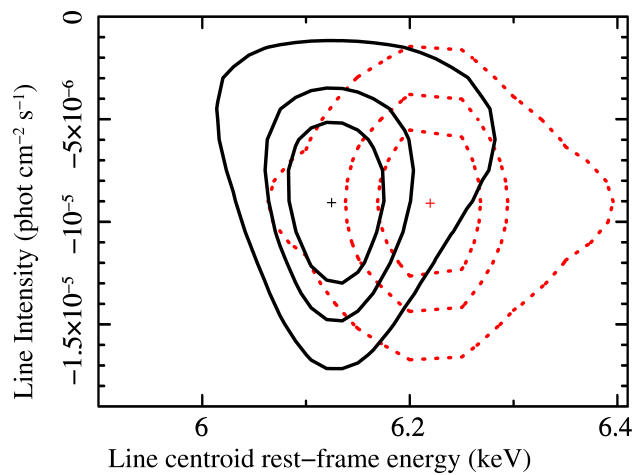}
   \caption{Confidence contours of 68, 90, and 99\% significance for the line centroid energy vs. its normalisation for the OBS1 (black solid lines) and for the OBS2 (red dotted lines) observations. The underlying continuum is Model A \texttt{xillver + tbnew\_pcf*(blackbody + power law).}}
              \label{FIGURE7}%
    \end{figure}

As the three models are statistically equivalent, we proceed with the analysis of this absorption feature using Model A (see e.g. the left columns of Table~\ref{TABLE:FIT} and Figure~\ref{FIGURE5}) as the baseline model, since it is the simplest and can be very quickly computed using modern workstations. 
  The line has an observed centroid energy $E_{line}^{OBS1}=6.05\pm{0.03}$ keV and $E_{line}^{OBS2}=6.13\pm{0.05}$ keV, corresponding to $E_{line}^{OBS1}=6.13\pm{0.03}$ keV and $E_{line}^{OBS2}=6.22\pm{0.04}$ keV in the source rest frame. The top panel of Figure~\ref{FIGURE7} reports the 68, 90, and 99\% confidence level contours for the rest-frame centroid energy of the line versus its intensity for the OBS1 (black solid lines) and OBS2 (red dotted lines) observations.
  The line equivalent width in the source rest frame is EW$^{OBS1}=-26^{-8}_{+7}$ eV and EW$^{OBS2}=-29^{-11}_{+9}$ eV.
 
 We checked for the presence of the absorption feature in the MOS data; albeit the MOS data were heavily piled-up, narrow spectral features should barely be affected by pile-up. We did not find strong evidence for the presence of such a feature, however, the photon statistics of the MOS data at energies $E\gtrsim 6$ keV is so low that the presence of the feature cannot be disproved either.
 
To better determine the statistical significance of the line, we therefore performed Monte Carlo simulations, following the method outlined in \citet{2006MNRAS.366..115M}.
As our null hypothesis we used the best-fitting parameters of Model A, without the inclusion of the redshifted absorption line.
Using the \texttt{fakeit} command within \texttt{Xspec}, single+double event spectra were simulated for OBS1 and OBS2 simultaneously, using the same exposure time as in the two XMM-Newton observations, and subtracting the corresponding backgrounds.
Such spectra were then fitted with Model A, obtaining new best-fitting parameters that provided a new and refined null hypothesis model, differing from the original only because of Poissonian statistics.
From this, refined model spectra were again simulated simultaneously for OBS1 and OBS2, using the same exposure time as in the two XMM-Newton observations, and subtracting the corresponding backgrounds; the simulated spectra were then fitted with Model A and the resulting $\chi^2$ was recorded.
A narrow Gaussian absorption line was then added to the model, with the normalisation free to vary between OBS1 and OBS2 and the centroid energy varying from $5$ to $8$ keV in steps of 100 eV; the maximum difference in chi-square ($\Delta\chi^2$) was then recorded. 

The steps above were repeated 1000 times and an equal or greater $\Delta\chi^2$ compared to our observed one ($\Delta\chi^2$=21) was found to randomly happen five times. This corresponds to a statistical significance of 99.5\%, equivalent to $\sim 2.81\sigma$. 

 The observed line centroid energy falls in a region that does not have known atomic transitions \citep[see e.g.][]{1997A&AS..125..149D}. The only possible physical interpretation is to associate it with \textit{redshifted} absorption, likely from highly ionised iron. 

 \begin{table}
\caption{Fit statistics for the highly ionised redshifted absorber.\label{TABLE:ABS}}
\centering
\begin{tabular}{llc}
\hline\hline
$^{(a)\,}N_H$ & $3.5\pm{1.3} \times 10^{23}$ & cm$^{-2}$\\
$^{(b)\,}\log\xi$ & $4.0\pm{0.2}$ & erg cm s$^{-1}$\\ 
$^{(c)\,}\upsilon_{shift}$ & $+3.6\pm{0.2}\times 10^4$ & km s$^{-1}$\\
\hline
\end{tabular}
\tablefoot{Notes: (a) Absorber column density; (b) Logarithm of the absorber ionisation parameter; (c) Absorber velocity shift.}
\end{table}

 \subsubsection{A physical model}    
In order to test the scenario described in the previous Section with a more physically realistic model, we removed the Gaussian absorption line and added a layer of ionised gas modelled with \texttt{Xabs} within \texttt{SPEX}, focusing on the long OBS1 observation.
\texttt{Xabs} is an absorption model that, given an appropriate ionising continuum, makes use of \texttt{Cloudy} to self-consistently compute the ionisation balance of the gas.
We used the input ionising continuum as given by the OM and EPIC-pn simultaneous measurements applied to Model A, corrected for Galactic absorption.

\begin{figure}
   \centering
  \includegraphics[width=9 cm]{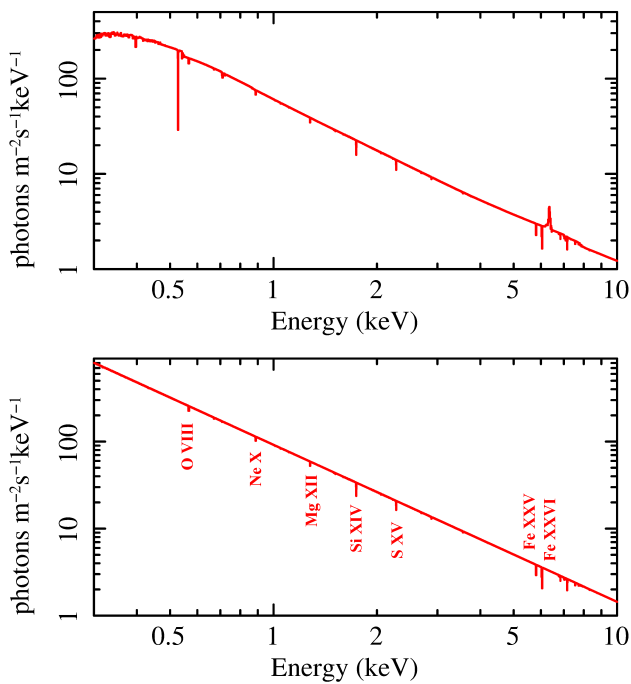}
   \caption{Top panel: total best-fitting theoretical model, i.e. a power law plus blackbody plus reflected emission that is absorbed by Galactic absorption, an intrinsic neutral partially covering absorber, and an intrinsic highly ionised redshifted absorber; the strong feature at $\sim 0.5$ keV is due to Galactic oxygen absorption. Bottom panel: the effect of the highly ionised redshifted absorber on the power law emission alone is plotted; the strongest absorption lines are labelled. }
              \label{FIGURE8}%
    \end{figure}

The result is an improvement of the fit statistics by $\Delta\chi^2/\Delta\nu=20/3$, where the free parameters are the absorber column density $N_H\sim  3\times 10^{23}$ cm$^{-2}$, ionisation parameter $\log\xi\sim 4$, and velocity shift $\upsilon_{shift}\sim 0.1c$. Model parameters along with their errors are reported in Table~\ref{TABLE:ABS}, while the theoretical best-fitting model spectrum is plotted in the top panel of Figure~\ref{FIGURE8}. The effect of the redshifted ionised absorber alone is plotted in the bottom panel of Figure~\ref{FIGURE8}: the most intense absorption features are due to highly ionised, redshifted Fe.
 
\begin{figure}
   \centering
  \includegraphics[width=8cm]{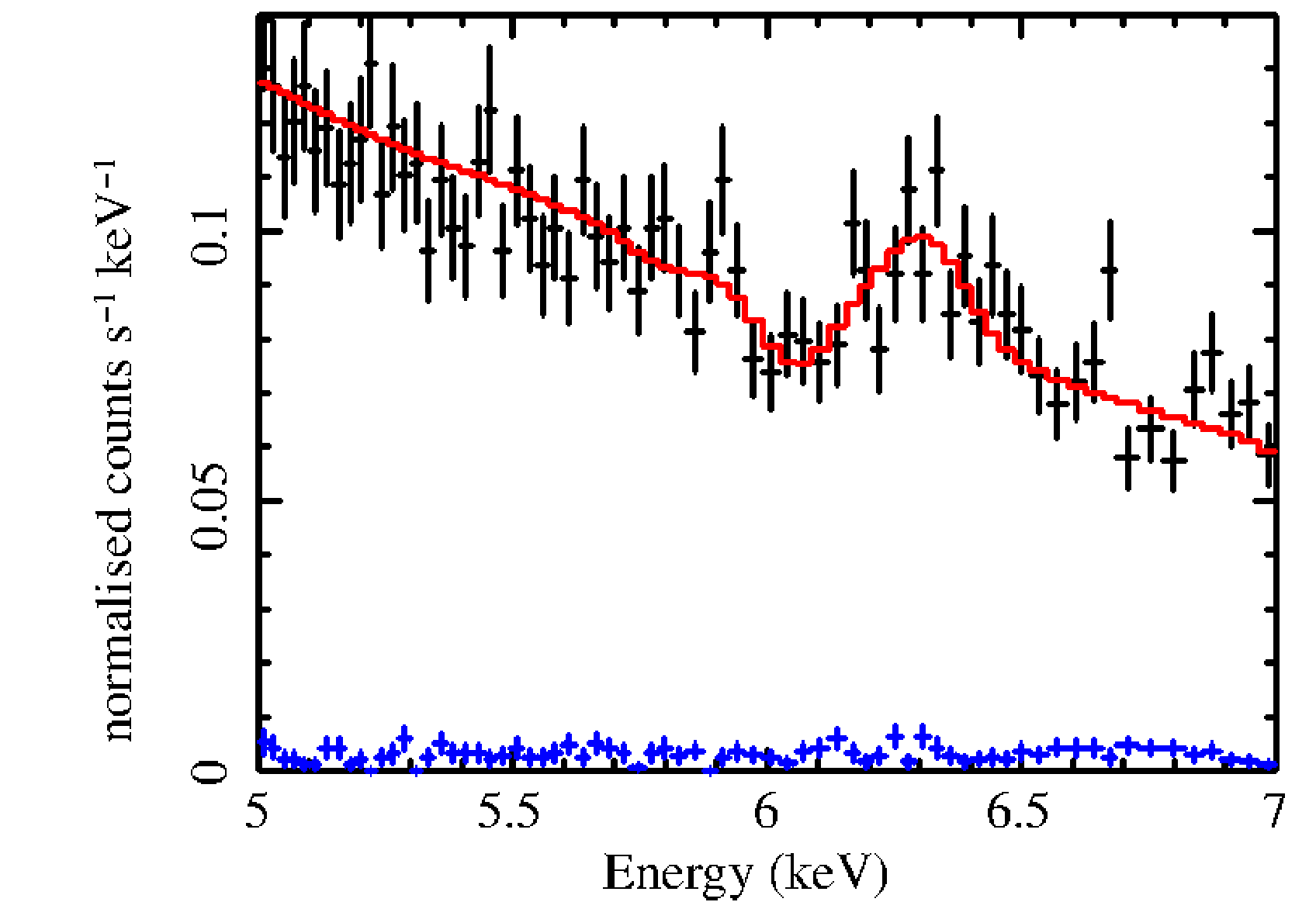}
  \includegraphics[width=8 cm]{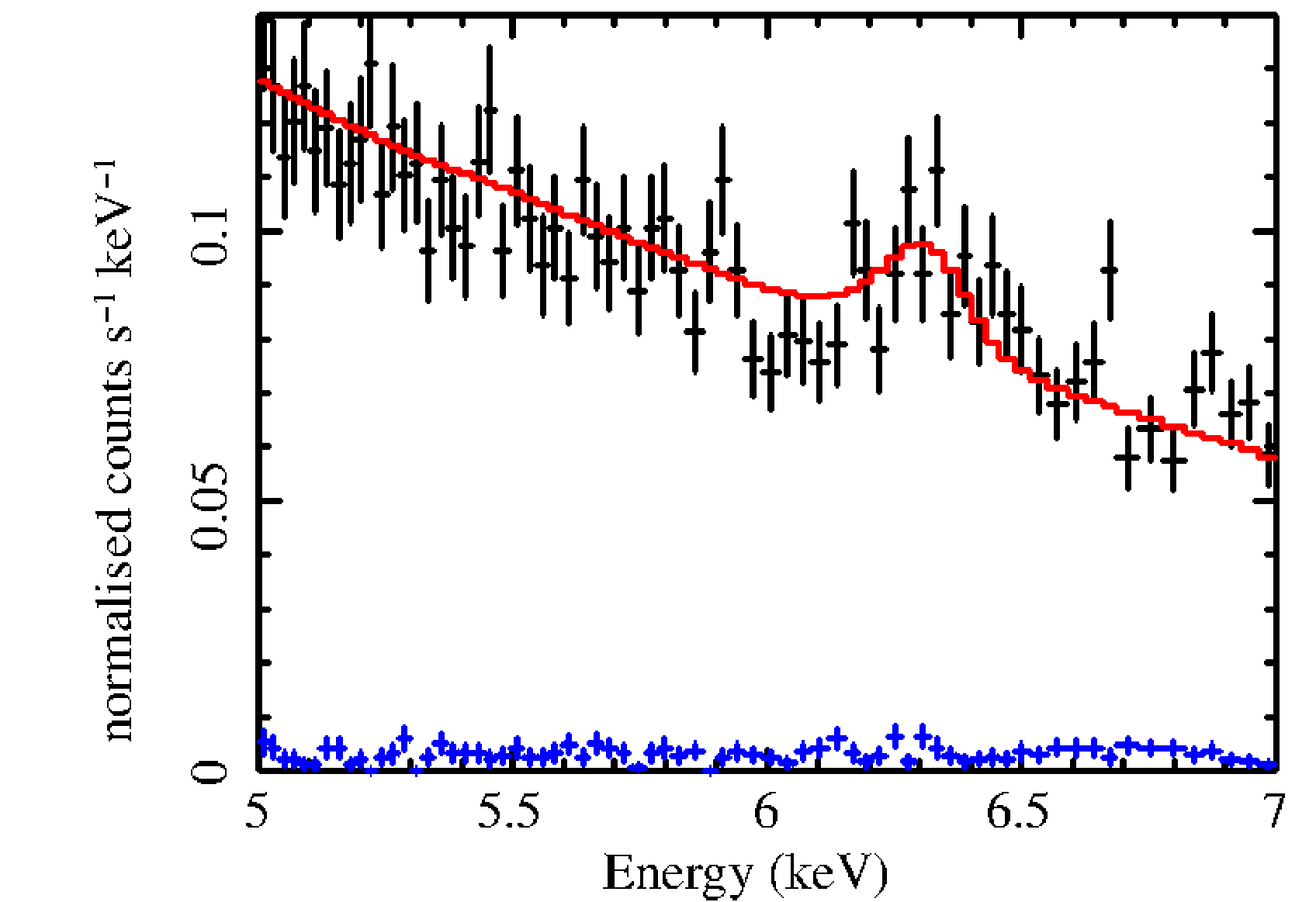}  
   \caption{Zoom on the Fe K region for the best-fitting model (top panel) and for the case where the data were fitted with the model without the redshifted absorber (bottom panel).
   Background-subtracted data points are plotted in black, the model is plotted in red, and the background is plotted in blue. }
              \label{FIGURE9}%
    \end{figure}

In Figure~\ref{FIGURE9} we show a zoom on the Fe K region showing the model (red line) superimposed on the data points (black) for the best-fitting model (top panel) and then we show the model with the highly ionised redshifted absorber removed (bottom panel). For comparison, the background is plotted in blue. It is clear that there is negligible contribution of the background to the observed data, and the absorption feature cannot be due to an incorrect subtraction of the background. 

  \subsection{The RGS Ssectrum}
\begin{figure}
   \centering
  \includegraphics[width=6.5 cm, angle=-90]{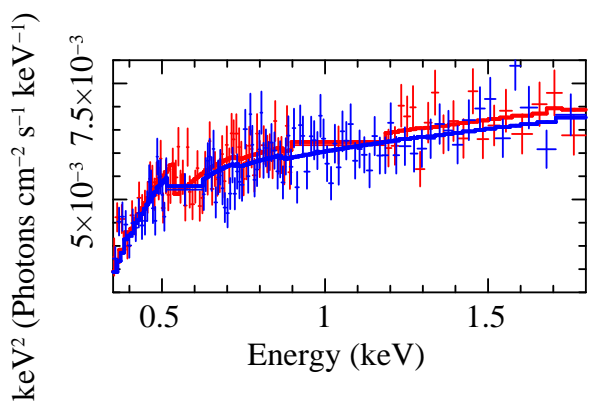}
   \caption{RGS1 (red) and RGS2 (blue) spectra unfolded against the best-fitting EPIC-pn Model A (thick solid lines), where the data have been visually rebinned to $15\sigma$ significance.}
              \label{FIGURE11}%
    \end{figure}

Blue shifted ionised absorption (the so-called warm absorber) is observed in about a half of type 1 AGN \citep[e.g.][]{2007MNRAS.379.1359M}.
The RGS spectra of OBS1 and OBS2 of NGC 2617 were analysed to look for the presence of warm absorbers and emission, but no evidence was found. 
The RGS spectra are shown in Figure~\ref{FIGURE11} unfolded against the EPIC-pn best-fitting Model A, where only the continuum normalisations were left free to vary.
    
However, our best-fitting models predict a few atomic features to be present in the NGC 2617 soft X-ray spectrum. 
Models B and C predict a strong emission line due to O~VII at $E\sim 560$ eV (see the right panels of Figure~\ref{FIGURE5}). 
While no evidence of discrete emission lines is found in the RGS spectrum, given the strength of the predicted emission line and the flux level of the source, the presence of such feature cannot be ruled out.  
Also the redshifted absorber model predicts a few absorption lines in the RGS band (see the bottom panel of Figure~\ref{FIGURE8}). In particular, significant absorption with ionic column densities $> 10^{16}$ cm$^{-2}$ is predicted for the H-like ions of oxygen, neon, magnesium, and silicon. Among these transitions, the O VIII falls in a region contaminated by the presence of bad pixels, while the Ne X, Mg XII, and Si XIV lines fall in a region where the degradation of both resolution and effective area of the RGS makes it difficult to assess the presence of such features. 
Thus from the RGS data we cannot discard the presence of these absorption features, and we conclude that there is no tension between the data and our absorption model.

\section{Discussion}\label{Section:DISCU}

We presented a comprehensive analysis of two XMM-Newton observations of the changing-look Seyfert 1.8 to 1.0 galaxy NGC 2617 \citep{2014ApJ...788...48S}.
The observations were performed one month apart; within this time interval the source flux roughly doubled, while the spectral shape did not vary dramatically; basically only the photon index of the primary power law emission steepened by $\Delta\Gamma\sim 0.1$. 
Within the observations, the source flux varied by less than 8\% in the $0.3-10$ keV energy range (Fig.~\ref{FIGURE1} and Fig.~\ref{FIGURE2}). 

\subsection{The AGN internal structure}
High-resolution X-ray spectroscopy performed with the RGS and focused on the soft X-ray band ($0.35-1.8$ keV) revealed the source to be bare, displaying no strong signs of ionised blueshifted absorption from a warm absorber. 
Broadband spectroscopy performed with the EPIC-pn over the $0.3-10$ keV band revealed the source to appear, to the first order, as a type 1 AGN in the X-ray band. 

However, there are deviations from a simple power law emission, with a slight excess of soft X-ray emission and with significant structures in the Fe K band accompanied by a substantial hardening at the highest energies probed by XMM-Newton.
Independent of the continuum adopted (Section~\ref{SEC:CONTINUUM} and Table~\ref{TABLE:FIT}), the source spectrum requires the presence of $\sim 10^{23}$ cm$^{-2}$ of neutral gas partially covering ($C_f\sim 0.33-0.45$, depending on the continuum model) the primary source.
This layer of partially covering gas helps reproduce the spectral hardening of NGC 2617 at the highest energies probed here. If such a spectral hardening was due to reflection alone, a much higher flux for the iron emission line(s) would be predicted by the theoretical model. However, these models cannot resemble the data, even by allowing the Fe abundance to drop to half of the solar value.

The iron emission line is indeed modest, with an EW$\sim 50$ eV. Its flux increases significantly between the two exposures, following the overall increase in flux of NGC 2617.
This suggests that the gas responsible for such emission is located not farther away than one light-month (that is the time elapsed between the two observations) from the continuum emission source. This corresponds roughly to $1.3\times 10^4$ $r_g$ for a black hole mass of $M_{BH}\sim 4\times 10^7$ $M_{\odot}$ as estimated by \citet{2014ApJ...788...48S} from the width of the broad optical emission lines in NGC 2617, and in agreement with the values expected from the observed Fvar \citep[Section~\ref{SEC:TIMING}][]{2012A&A...542A..83P}.
This is much smaller than the location of the putative ``torus'' that surrounds the AGN central engine, but rather corresponds to the broad line region/accretion disk spatial scale \citep[see e.g.][]{1993ARA&A..31..473A}. The normalisation of the reflected emission component also follows the increase in flux of the source, all in all suggesting that we are observing reflection from material closer in than the ``parsec scale torus''.

As for the intrinsic underlying continuum, the limited bandpass of XMM-Newton makes it hard to distinguish between a double partial covering absorption (Section \ref{PCOV}) and a double Comptonisation (Section \ref{COMP}) scenario, the two models being statistically equivalent. 
Even adding the INTEGRAL IBIS/ISGRI data up to $E\sim 100$ keV, the fit statistics is equivalent for the different scenarios.
The main difference in terms of physical parameters are a photon index $\Gamma$ that is steeper by a factor of $\sim 0.2-0.25$ and a higher intrinsic $0.3-10$ keV luminosity by a factor of $\sim 20\%$ in the case of the double partial covering absorption scenario compared to the double Comptonisation scenario. 

The double Comptonisation scenario has been recently found to be able to reproduce the broadband spectra of several AGN \citep[see e.g.][and references therein]{2013A&A...549A..73P, 2014A&A...563A..95D, 2015A&A...575A..22M, 2015A&A...577A...8G}.
 In this scenario, the ``warm corona'' emission would dominate at $E<2 $ keV, thus accounting for the soft excess. However, this model is not able to fit the OM data (Section~\ref{SEC:EXTENSION}). 

The \texttt{optxagn} model gives Eddington ratios $L/L_{Edd}\sim 0.035$ and $0.112$ for OBS1 and OBS2, respectively. Given the black hole mass of NGC 2617 \citep[$4\times10^7 M_{\odot}$,][]{2014ApJ...788...48S}, the Eddington luminosity is $L_{Edd}\sim 5\times 10^{45}$ erg s$^{-1}$ and this translates to bolometric luminosity of $L_{Bol}\sim 1.8-5.6\times 10^{44}$ erg s$^{-1}$ for OBS1 and OBS2. Compared with the measured (de-absorbed) $2-10$ keV luminosity $L_{2-10}\sim 1.4-2.9\times 10^{43}$ erg s$^{-1}$ gives a $2-10$ keV bolometric correction $\kappa_{2-10}=L_{Bol}/L_{2-10}\sim 13-20$, in very good agreement with the behaviour displayed by the 'average' type 1 AGN at such mass accretion rates \citep{2012MNRAS.425..623L}. The double partial covering scenario would give very similar bolometric corrections for Eddington ratios that are $\sim 30\%$ higher.

Also the short-term variability properties of the source display behaviours that are typical of a bare type 1 AGN. 
The measured Fvar on timescales of $20-3$ ks is consistent with the excess variance-BH mass correlation reported in \citet{2012A&A...542A..83P}, characterising type 1 AGN. 
The tentative detection of a soft X-ray lag might be indicative of reprocessing in the innermost regions of an optically thick accretion disk, as commonly observed in type 1 AGN \citep{2013MNRAS.431.2441D}. 
The spectral-timing response of complex absorbers to variations in the ionising flux of the central source was studied in \citet{2016arXiv160701065S}. However, according to our model, the highly ionised redshifted absorber does not imprint strong features in the soft X-ray band (Fig.~\ref{FIGURE8}). The lags produced by transient absorption phenomena such as eclipsing clouds (e.g. from the partial covering gas included in our models) have been studied in the literature \citep{2014MNRAS.442.2456G}. In particular, a soft lag is expected as a result of an increase of hard X-ray flux followed by an increase of soft X-ray flux as an optically thick cloud moves out of the line of sight \citep{2015MNRAS.446..737K}. However, the best-fit parameters of the partial coverer are consistent with being constant between the two observations and we cannot put constraints on their possible variations within the single exposures. It is possible that the lag is produced by reflected light, in this case the absorber/reflector should be much closer in than the neutral reflection component that we found in our spectral analysis. 

In any case,  the amplitude of the lag corresponds to a distance of a few $r_g$ from the primary X-ray source, assuming a BH mass of $4\times 10^7 M_{\odot}$.
All in all, these variability properties on a short timescale point to a genuine type 1 AGN nature for the nucleus of NGC 2617.

\subsection{An AGN accretion event}

The most striking result of our analysis is the detection of a redshifted iron absorption line in both observations. 
While blueshifted absorption lines are commonly detected in the X-ray spectra of type 1 AGN, the occurrence of redshifted absorption is much rarer.
Blueshifted iron absorption lines associated with powerful nuclear winds are indeed observed in about a half of type 1 AGN \citep[e.g.][]{2010A&A...521A..57T}. 
Such winds are of great importance as they are possibly able to exert a ``negative'' feedback on the host galaxy (Silk \& Rees 1998).

On the other hand, the detection of \textit{redshifted} iron X-ray absorption has only been reported for a few sources, namely NGC 3516 \citep{1999ApJ...523L..17N}, E1821+643 \citep{2005ApJ...623..112Y}, Mrk 509 \citep{2005A&A...442..461D}, PG 1211+143 \citep{2005ApJ...633L..81R}, Q 0056-363 \citep{2005A&A...435..857M}, Mrk 335 \citep{2007MNRAS.374..237L}, and CID-42 \citep{2010ApJ...717..209C}. 
The general interpretation for the presence of such redshifted X-ray absorption was an inflow of matter towards the central SMBH (i.e. an AGN accretion event) and/or gravitational redshift, except for CID-42 where, because of the peculiar nature of the source, a recoiling SMBH scenario was discussed.
In every case the detection consisted of a single absorption line associated with highly ionised iron, and the redshifted absorption line was not detected again in any single case where the source was re-observed. 

Conversely, in NGC 2617 a redshifted iron absorption line is clearly present in both XMM-Newton observations, performed one month apart in 2013 (see Fig.~\ref{FIGURE7} and \ref{FIGURE9}).
By means of extensive Monte Carlo simulations we assessed the statistical significance of the line to be $99.5\%$, which is equivalent to $\sim 2.8\sigma$. 
We note how, given the low-resolution X-ray spectral regime we are dealing with, such significance is among the highest ever measured in the Fe K band of an AGN: as an example, of the AGN sample studied by \citet{2010A&A...521A..57T}, only 6 out of 22 detected absorption lines have a Monte Carlo significance $\geq 99.5\%$. 
Given the measured strength of the absorption lines and their errors, our result also falls into the ``safe'' part of the EW$/\Delta(\rm{EW})$ diagram presented by \citet{2008MNRAS.390..421V}. 

When modelled with a physically self-consistent absorption model that takes the NGC 2617 spectral energy distribution into account, the absorber is found to have a high ionisation state, $\log\xi\sim 4$, a large hydrogen-equivalent column density $N_H\sim 3\times  10^{23}$ cm$^{-2}$, and a velocity shift $\upsilon_{shift}\sim 36,000$ km s$^{-1}$ (see Table~\ref{TABLE:ABS} and Fig.~\ref{FIGURE8}). Several scenarios are open to interpret the redshifted iron absorption observed in NGC 2617. We discuss three of these scenarios below: matter falling towards the central SMBH, failed wind/aborted jets, and gravitational redshift effects.

\subsubsection{Matter falling towards the central SMBH}
Given the physical quantities and timescales involved in this study, the ``simple'' yet intriguing infall of matter towards the central SMBH may hold.
Assuming a very simple free-fall motion, the time taken for the absorber to free fall into the SMBH can be written as
$$
\tau_{ff} \sim 250 \left( \frac{M_{BH}}{M_{\odot}}\right) \left( \frac{\upsilon_{shift}}{1,000\, \rm{km\, s^{-1}}} \right)^{-3}\:\rm{s}.
$$
Plugging in the quantities relevant to NGC 2617, one finds $\tau_{ff}\sim 230$ ks, which is equivalent to about 2.5 days.
The associated radial scale is about $8\times 10^{14}$ cm, corresponding to $\sim 130 r_g$ in the case of NGC 2617.
The last optical spectrum taken of NGC 2617 prior to its transition to a type 1 AGN is from late December 2003; therefore the transition must have occurred sometime during the ten years elapsed until the detected outburst in 2013. 
With the observed velocity a range of radial distances $10^{3-5} r_g$ could be covered in $\sim 10$ years. 
Again, the distances implied are on sub-parsec scales, i.e. on the scales typical of an accretion disk \citep[see e.g.][]{2002apa..book.....F}.

The adoption of a free-fall motion is an obvious oversimplification for a complex situation like the accretion of material around a SMBH, and can be treated as the shortest possible timescale for matter to accrete onto the SMBH, while the viscous timescale in an accretion disk is likely to be hundreds of times longer.
However, in the chaotic accretion scenario \citep[e.g.][]{2006MNRAS.373L..90K} where individual ``blobs/clouds'' of matter get accreted by the SMBH, we are seeing accretion from outside the plane of the disk, therefore the disk viscosity timescale is not relevant.
In the non-spherical hydrodynamical simulations of an accreting SMBH performed by \citet{2012MNRAS.424..728B}, the coexistence of a cold, clumpy, and fast accreting phase and of a hot buoyant phase is predicted. This is in agreement with our detection of both highly ionised absorption and cold partially covering absorption.


\subsubsection{Failed wind/aborted jet}
Another possible explanation for the observed redshifted absorption lines is of dynamical origin, as in a failed inner disk wind/aborted jets \citet{2004ApJ...616..688P} or an ``aborted jets'' scenario \citet{2004A&A...413..535G}. 

In the failed disk wind scenario, the material responsible for absorption is part of a line-driven wind that is trying to escape the system, but gets over-ionised by the strong continuum radiation, loses momentum, and eventually falls back towards the plane of the accretion disk.
In this scenario, we expect the timescales of variations to be short, of the order of the dynamical/free-fall timescale. The observation of the same energy/redshift for the absorption line during OBS1 and OBS2, which are spaced by much more time than the free-fall timescale, could make this scenario a bit contrived, however, the limited spectral resolution of the EPIC-pn prevents us from discarding this scenario on the basis of this argument.
In the ``aborted jets'' scenario, a blob of ionised matter is launched from the vicinity of the central SMBH with a velocity that is insufficient to escape the system, therefore it falls back towards the centre under the gravitational pull of the SMBH. 
In both scenarios an outflowing (i.e. blueshifted) component would also be expected; there are hints of blueshifted absorption in the spectrum of NGC 2617, but unfortunately the low statistics at high energies does not allow us to assess the presence of such features at more than the $2\sigma$ level.

\subsubsection{Gravitational redshift effects}
 If the absorber is located very close to the SMBH, general relativistic effects such as gravitational redshift may play an important role \citep[e.g.][]{1973grav.book.....M}. 
 The formula for the gravitational redshift at a given distance $r$ from the central SMBH in units of gravitational radii is
 $$z_{grav} =\left[ 1/\sqrt{(1-2r_g/r)} \right] - 1.$$
Defining $\zeta = E_{observed}/E_{intrinsic}$, we can rewrite this equation as:
$$r/r_g = 2/(1-\zeta^2).$$
For $E_{observed} \sim 6.1$ keV and $E_{intrinsic} \sim 6.4$ keV (zero ionisation), we get $r \sim 20 r_g$. For $E_{intrinsic} \sim 6.97$ keV (highest ionisation), we only obtain $r\sim 8.5 r_g$.

However, the estimation above is a static solution, while the absorber is most likely in motion. 
If we assume a simple cloud orbiting with a Keplerian velocity, then a special relativistic transverse Doppler shift needs to be taken into account. We can estimate the radial distance using an approximation derived by \citet{2005A&A...441..855P} for a simple non-rotating black hole. 
The relation for the frequency shift at the line of sight is
$$\zeta = \sqrt{[(r/r_g-3)/r/r_g]},$$
i.e.
$$r/r_g = 3/(1-\zeta^2).$$

For $E_{intrinsic} \sim 6.4$ keV, we get $r \sim 33 r_g$, for $E_{intrinsic} \sim 6.97$ keV, we get $r\sim 13 r_g$.

We can estimate the orbital time of the cloud at such distance as
$$\tau_{orb} \sim 31 (r/r_g)^3/2 (M/10^6M_{\odot})\, \textrm{s},$$
which gives $\tau_{orb}\sim 240$ ks for $E_{intrinsic} \sim 6.4$ keV and $\tau_{orb}\sim 60$ ks for $E_{intrinsic} \sim 6.97$ keV, respectively.
At such vicinity to a black hole, any orbiting cloud would be sheared by tidal forces and would form an extended structure along the orbit and the obscuration can therefore take place during a significant part of the orbit.
We might therefore expect to detect variability of the absorption line within and between the two observations. 
There are hints of variability of the absorption line within the long OBS1, but by dividing the observations in time slices the statistics become too low to assess such variations with a reasonable significance. 
There are also hints of variability of the absorption line between the two observations in energy and intensity, but such variations are again statistically not significant (Figure ~\ref{FIGURE7}). All in all, the variability of the redshifted iron absorption line will remain speculative until a longer, deeper X-ray observation of NGC 2617 will possibly allow time slices to be properly analysed.
If gravitational redshift is at work we should also expect to observe an asymmetric, distorted absorption line profile. The spectral resolution of the EPIC-pn is too low to assess the shape of the absorption trough, an exercise that could be carried out with a higher spectral resolution instrument.

If the absorber is an outflow rather than an orbiting cloud, the situation may be more complicated as the local physics and Doppler shift due to radial velocity will be important in shaping the exact spectral line profile. The case of an expanding shell around a compact object was studied by \citet{2009MNRAS.393.1433D} in pseudo-Newtonian approximation. For some parameters, they indeed obtained a significant redshift of the line, usually together with a blueshifted component as well.

Also, the estimated distance of the absorber, if the line frequency shift is mainly due to the gravitational redshift, is comparable to the estimated extension of the warm corona measured by the \texttt{optxagnf} model in the double Comptonisation scenario. 
This suggests that the absorbing gas may be a part of this geometrically thick corona, possibly being its highly ionised outer layer, and both effects, the gravitational redshift and the inflow to the black hole, may play a role together.

   \section{Conclusions}\label{Section:CONCLU}
We presented the comprehensive analysis of two XMM-Newton observations of the changing-look AGN in NGC 2617, which recently switched from being a Seyfert 1.8 to be a Seyfert 1.0, and at the same time underwent a strong broadband flux increase: in particular, its X-ray flux increased by one order of magnitude with respect to archival measurements \citep{2014ApJ...788...48S}. The two observations were performed in 2013 and were separated by one month. 
During this time the X-ray flux of the source doubled, while the spectral shape did not dramatically change.\\

\noindent The main results of our work can be summarised as follows:
\begin{itemize}
\item To the first order, at X-ray wavelengths NGC 2617 appears as a type 1 AGN, being dominated by a power law-like continuum emission with no signs of absorption fully covering the source.\\
\item A modest reflection component is detected, with an associated narrow iron Fe K emission line of equivalent width EW$\,\sim 50$ eV.\\
\item The NGC 2617 underlying continuum can be modelled in a statistically equivalent way with a power law plus a phenomenological soft blackbody emission, with a power law partially covered by a neutral absorber, or a double-Comptonisation model. In either case, a further high column ($N_H\sim 10^{23}$ cm$^{-2}$) of neutral gas only partially covering ($C_f\sim 0.2-0.4$) the continuum source is required by the data.\\
\item Once we account for partially covering absorption, the intrinsic power law emission slope of NGC 2617 is typical of type 1 AGN, $\Gamma\sim 1.9$, and steepens by $\Delta\Gamma\sim 0.1$ when going from the low (OBS1) to the high (OBS2) flux level: this steepening when brighter is also typical of type 1 AGN.\\
\item NGC 2617 varies by less than  $8\%$ within the single observations.\\
\item The short-term variability properties (i.e. the measured Fvar and the tentative soft X-ray lag) are typical of a type 1 AGN with a black hole mass of $M_{BH}\sim 4\times 10^7\, M_{\odot}$.\\
\item The RGS data showed the lack of any strong warm absorption/emission in the soft X-ray spectrum of NGC 2617, possibly making it a ``bare'' Seyfert 1 galaxy.\\
\item Redshifted X-ray absorption is detected in both observations, with a global significance of $99.5\%$. The absorption is compatible with an equivalent column density of $N_H\sim 10^{23}$ cm$^{-2}$ of highly ionised iron that is redshifted by $\sim 35,000$ km s$^{-1}$.\\
\item Possible interpretations for such a redshifted, high-velocity absorber include a failed disk wind/aborted jets, gravitational redshift, and/or an infall of matter towards the central supermassive black hole of NGC 2617. All these scenarios are compatible with the observed properties of NGC 2617.\\
\end{itemize}

Overall, the X-ray properties of NGC 2617 are indicative of a genuine Seyfert 1 nucleus. 
Most strikingly, the detection of persistent redshifted X-ray absorption associated with highly ionised gas opens the way to starting to directly probe the inner accretion flow around supermassive black holes.

\begin{acknowledgements}
The Space Research Organisation of the Netherlands (SRON) is supported financially by NWO, the Netherlands Organisation for
Scientific Research. 
This work is mainly based on observations obtained with XMM-Newton, an ESA science mission with instruments and contributions directly funded by ESA Member States and NASA.
This work is also partially based on observations obtained with INTEGRAL, an ESA project with instruments and science data centre funded by ESA member states (especially the PI countries: Denmark, France, Germany, Italy, Switzerland, Spain), and with the participation of Russia and the USA.
MG would like to thank Norbert Schartel for scheduling the DDT and granting the ToO XMM-Newton observations of NGC 2617, Chris Done for discussions about the \texttt{optxagn} model and the AGN structure, and Javier Garc\'ia for discussions about \texttt{xillver} and other reflection models. 
We thank the anonymous referee for providing a constructive and careful report that allowed us to significantly improve the quality of the manuscript.
This research has made use of the NASA/IPAC Extragalactic Database (NED), which is operated by the Jet Propulsion Laboratory, California Institute of Technology, under contract with the National Aeronautics and Space Administration.
MG was partially supported by an ESA Research Fellowship in Space Science.
JS acknowledges support from the Grant Agency of the Czech Republic within the project 14-20970P and the Czech-U.S. collaboration grant LH 14049.
Support for JLP is in part provided by FONDECYT through the grant 1151445 and by the Ministry of Economy, Development, and Tourism's Millennium Science Initiative through grant IC120009, awarded to The Millennium Institute of Astrophysics, MAS.
BS was supported by NASA through Hubble Fellowship grant HF-51348.001 awarded by the Space Telescope Science Institute, which is operated by the Association of Universities for Research in Astronomy, Inc., for NASA, under contract NAS 5-26555.   \end{acknowledgements}
    
\bibliographystyle{aa} 
\bibliography{mybib}

\end{document}